\documentclass[final,12pt]{elsarticle}

\usepackage{amssymb}
\usepackage{lineno}

\newcommand{\degr}{$^{\circ}$} 

\begin{document}

\begin{frontmatter}


\title{Concept of round non-flat thin film solar cells and their power conversion efficiency calculation}


\author{Jabbar Ganji}

\address{Dept. of Electrical Engineering, Mahshahr Branch, Islamic Azad University, Mahshahr, Iran}

\begin{abstract}
Thin-film solar cells that are considered as the second generation of solar cells are known for their low cost and acceptable efficiency. In this technology, semiconductor layers with a thickness of micrometer are deposited on thick enough substrates to maintain physical consistency. The relatively low processing temperature helps use substrates of different materials. Compared with crystalline solar cells, which are mainly made up of rigid flat plates, it is also possible to make thin-film cells on flexible or non-flat substrates. In this study, a method was first proposed to calculate the efficiency of such cells without the need for 3D simulation, and then it is investigated using non-flat conical and paraboloid substrates as a novel method to enhance the light trapping. As a result, a significant increase in the efficiency of the studied non-flat cells was observed and reported in comparison with the flat cells. In addition, the paraboloid shape shows a better performance than that of the conical, to use as the cell's substrate.
\end{abstract}


\end{frontmatter}



\section{Introduction}
\begin{figure}
	\begin{center}
		\resizebox{0.45\textwidth}{!}
		{\includegraphics{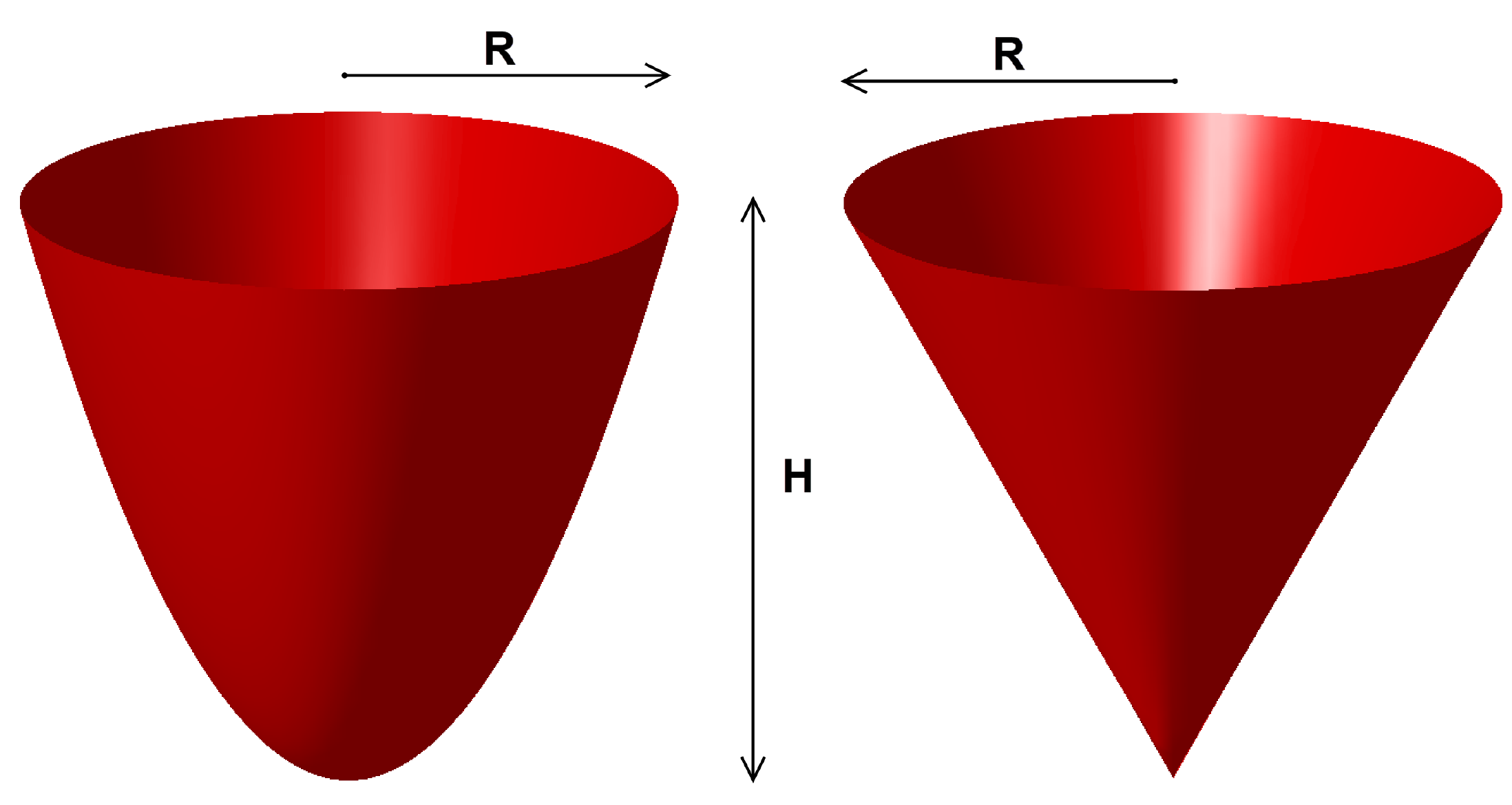} }
		\caption{\label{fig1}Two substrate shapes used in this study.}
	\end{center}
\end{figure}

Using different light-trapping methods, the optical length can be multiplied by several times of cell's physical thickness to increase the chance of photon absorption and, consequently, the power conversion efficiency ($\eta$) \cite{RN1,RN2}. Common methods of light trapping include: the use of anti-reflection coating \cite{RN3,RN4}, layers surface texturing \cite{RN5,RN6}, the addition of intermediate reflective layers \cite{RN7,RN8}, the usage of various majorly metallic nanoparticles in the absorber layers \cite{RN9,RN10} and using photonic crystals and plasmonic principles \cite{RN11,RN12,RN13}. \\
One of the other method of light trapping is to use the thin film solar cell with non-flat substrates especially macroscopic geometrical shapes \cite{RN13a}. Many references have been mostly suggested this idea to investigate the organic cells with folded (V-shaped) substrates \cite{RN13b,RN13c,RN13d,RN13e,RN13f}. Furthermore, there are limited reports on investigation of conical- \cite{RN13g} and cylindrical-shaped \cite{RN13h} substrates through the mentioned idea. This method of light trapping employs non-flat substrates in a way that a large portion of the reflected photons from a cell surface will re-emit to another section of the same cell in order to increase the absorption probability. The base of idea is to recycle the reflected and withdrawn photons of the cell.\\
In some thin-film deposition systems such as plasma-enhanced chemical vapor deposition (PECVD), in which the process temperature is low, it is possible to use a variety of flexible and cuttable materials as the substrate \cite{RN14}. The isotropic nature of chemical vapor deposition makes it possible to directly use the non-flat substrates. These two features help realize such an idea without the need to add expensive components or new sophisticated technology.\\
This paper studies two special substrate shapes: conical and paraboloid. The acuteness (a), which is defined as the ratio of the height to the span radius (H/R), is considered as the main parameter for both shapes (Fig. \ref{fig1}), and the power conversion efficiency is defined as the index parameter to evaluate the cells' merit. The two mentioned cells, are compared together in terms of the their efficiency and technological limitations. Towards this end, a simple method for optical and electrical analysis is developed.\\
The bulk dimensions of the cells (H and R) can be of any value theoretically and given the capabilities of recent deposition systems, choosing these two parameters on the order of centimeter is more preferable. In addition to direct use of the non-flat substrate in the deposition system, it is possible to make a thin-film cell with a flat and flexible substrate and then achieve conical geometry by cutting and removing a triangular section of it in order to realize the conical substrate.
\section{Simulation method}
\begin{figure}
	\begin{center}
		\resizebox{0.45\textwidth}{!}
		{\includegraphics{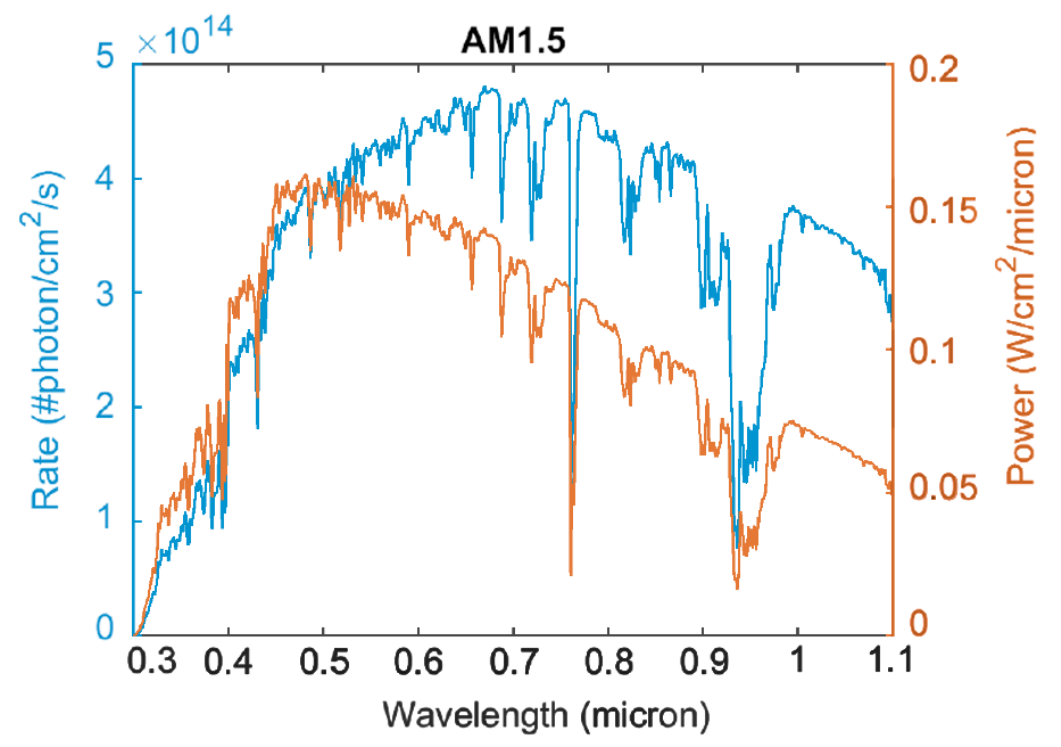} }
		\caption{\label{fig2}AM1.5 spectra.}
	\end{center}
\end{figure}

\begin{figure}
	\begin{center}
		\resizebox{0.35\textwidth}{!}
		{\includegraphics{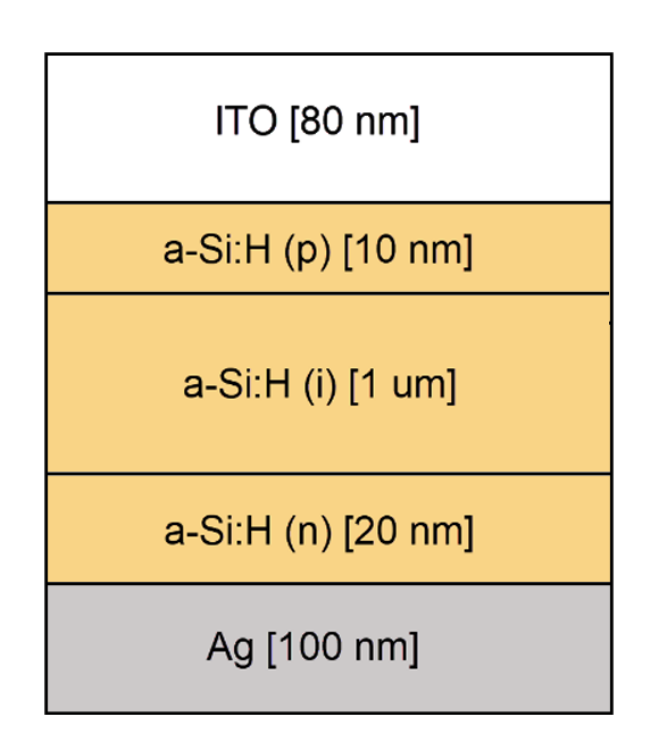} }
		\caption{\label{fig3}Structure of the base cell.}
	\end{center}
\end{figure}

\begin{figure}
	\begin{center}
		\resizebox{0.45\textwidth}{!}
		{\includegraphics{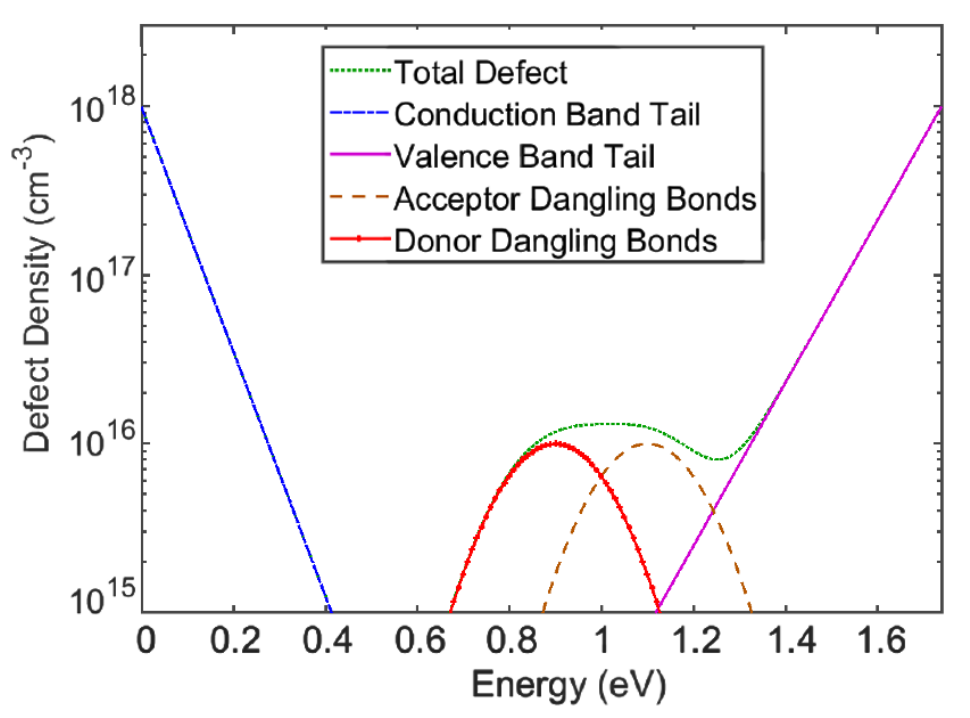} }
		\caption{\label{fig4}Defect density of states chart for i-layer.}
	\end{center}
\end{figure}

Three-dimensional simulation of a solar cell with non-flat substrate in centimeter dimension and micrometer thickness is very complicated, but considering the geometric order in the two substrates being studied and the ease of predicting the trajectory of radiation and reflection to their surfaces, it is possible to assume the non-flat substrate as many differential-sized flat parts having different angles to the horizon and to conduct a two-dimensional simulation on each piece. In this concept, the parameters of each stage of simulations are the angle of the incident light to the surface of the cell and its wavelength. A virtual flat thin-film cell with a certain area, called the ``base cell'', is defined as a sample for the differential-sized parts to be simulated. \\
The ray-tracing method is employed for optical part of the simulation. It is assumed that each single photon with a certain wavelength that incidents to a given point of the main cell, faces to a 1$\times$1 micron flat square surface at the incident position. This differential size surface is in fact, the base cell, which is considered as the area element of the main non-flat cell.\\
Simulations are performed in three phases. The first phase is the parametric simulation of the base cell, with two parameters of the wavelength of the transmitted monochrome photons and the angle of their incidence with the cell surface. The goal of this phase is to create a database of all possible photon-entry-absorb-exit states in order to minimize the need for reworking in the overall simulation process. A certain number of monochrome photons with parametric wavelengths and angles is radiated to it and in each step, the ratio of the absorbed photons to the transmitted ones is calculated by a computer simulation. The wavelength of the light is changed from 300 nm to 1100 nm at 25 nm intervals, and the radiation angle is from 0\degr to 90\degr and is swept at 5\degr intervals. For the angles less than 5\degr and more than 85\degr, the distance is considered as 1\degr for accuracy purposes. The data obtained from this series of simulations are stored in a data table, in which rows and columns correspond respectively to the wavelength and angles of incident photons. Each the table cell contains both the ratio of the converted power resulting from the photon absorption calculations and the ratio of the non-absorbed photons to radiated ones. Interpolation is used once the specified wavelength or angle does not exist in the database.\\
The second phase of the simulation is related to conducting geometrical and trigonometric calculations on the non-flat substrate shape and the direction of the light beam. This step of simulation is full analytical and its purpose is to obtain a chain of incident-reflection angles after the entrance of a vertical beam into the cell until the last beam's exit the cell. In each incidence of light to the cell surface, a number of photons are absorbed and the rests are reflected. The reflected photons re-emit to another surface of the cell and the cycle continues until the last non-absorbed photons exit the cell. After this cycle, the number of absorbed photons and total converted energy is calculated. Each photon's energy is calculated as
\begin{equation} \label{equ1} 
E_{ph} =\frac{hc}{\lambda }  
\end{equation} 
Where $E_{ph}$, $h$, $c$ and $\lambda$ are the photon energy, Plank's constant, light speed and photon wavelength, respectively. Further details on how to conduct this phase are described in following sections depending on substrate shape.\\
The third phase of simulation is the synthesis of the results of two previous phases by applying AM1.5 spectra \cite{RN15} data. The frequency table of AM1.5 spectra is used to determine the light intensity of each wavelength interval. Fig. \ref{fig2} shows the frequency of photons in a 1000 Watt AM1.5 spectra and their energies. This phase aims to calculate the total efficiency of the non-flat cell under standard test condition.\\
Although, the suggested method is not the most precise way to optical simulation of a general solar cell, however, it relies on low complexity and fair speed. \\
The structure of the base cell is considered as ITO/a-Si:H(p)/a-Si:H(i)/a-Si:H(n)/Ag (Fig. \ref{fig3}). Utilizing silver back contact helps remove the light transmission component which causes all non-absorbed photons to be reflected. So that the chance of further incidents with the cell's surface increases. The doping concentration of p and n are selected as $1\times 10^{20}$. Table \ref{tbl1} shows the general characteristics of this cell's layers. Defect-pool model is used for analyzing this cell  \cite{RN16,RN17,RN18}. Table \ref{tbl2} shows the information of the layers of base cell in the defect-pool model. The absorber layer's defect density of states chart is shown in Fig. \ref{fig4}. The efficiency of this cell under a vertically radiated 1000 $\rm{W/m^2}$ AM1.5 light is calculated as 6.124\%. 
\begin{table*} [htbp]
\caption{\label{tbl1}Common parameters of p, i, n layers used in the simulation.}						
\begin{center}
\centering
\begin{tabular}
{lccc} \hline 
Parameter & Symbol & Unit & Value \\ \hline 
Dielectric Constant & $\epsilon_r$ & - & 11.9 \\ 
Electron Affinity & $\chi_e$ & eV & 3.9 \\ 
Mobility Bandgap & ${E}_{g}$ & eV & 1.74 \\ 
Optical Bandgap & ${E}_{g,opt}$ & eV & 1.64 \\ 
Effective Conduction Band Density of States & $N_C$ & cm${}^{-3}$ & $2.5\times 10^{20}$ \\ 
Effective Valence Band Density of States & $N_V$ & cm${}^{-3}$ & $2.5\times 10^{20}$ \\
Electron Mobility & $\mu_e$ & cm${}^{2}$/Vs & 5 \\
Hole Mobility & $\mu_h$ & cm${}^{2}$/Vs & 1 \\
Electron Thermal Velocity & $v_e$ & cm/s &  $10^7$ \\
Hole Thermal Velocity& $v_h$ & cm/s & $10^7$ \\ \hline 
\end{tabular}
\end{center}						
\end{table*}
\begin{table*} [htbp]
\caption{\label{tbl2}Defect parameters of layers used in the defect-pool model at the current study.}						
\begin{center}
\centering
\begin{tabular}
{lcccc} \hline 
Parameter & Unit & p & i & n \\ \hline 
\multicolumn{5}{l}{Conduction Band Tail} \\ 
\quad Electron Thermal Cross Section & cm${}^{2}$ & $10^{-16}$ & $10^{-16}$ & $10^{-16}$ \\ 
\quad Hole Thermal Cross Section & cm${}^{2}$ & $2\times 10^{-16}$ & $2\times 10^{-16}$ & $2\times 10^{-16}$ \\ 
\quad Total Trap Density & cm${}^{-3}$ & $10^{18}$ & $10^{18}$ & $10^{18}$ \\
\quad Urbach Energy & meV & 70 & 70 & 60 \\  
\multicolumn{5}{l}{Valence Band Tail} \\ 
\quad Electron Thermal Cross Section & cm${}^{2}$ & $10^{-16}$ & $10^{-16}$ & $10^{-16}$ \\ 
\quad Hole Thermal Cross Section & cm${}^{2}$ & $3\times 10^{-17}$ & $3\times 10^{-17}$ & $3\times 10^{-17}$ \\ 
\quad Total Trap Density & cm${}^{-3}$ & $10^{18}$ & $10^{18}$ & $10^{18}$ \\
\quad Urbach Energy & meV & 120 & 120 & 90 \\ 
\multicolumn{5}{l}{Dangling Band Acceptor} \\  
\quad Electron Thermal Cross Section & cm${}^{2}$ & $5\times 10^{-17}$ & $5\times 10^{-17}$ & $5\times 10^{-17}$ \\ 
\quad Hole Thermal Cross Section & cm${}^{2}$ & $10^{-15}$ & $10^{-15}$ & $10^{-15}$ \\
\quad Total Trap Density & cm${}^{-3}$ & $10^{16}$ & $10^{16}$ & $10^{16}$ \\
\quad Energy of Distribution & eV & 1.3 & 1.1 & 0.7 \\ 
\quad Characteristic Energy & eV & 0.2 & 0.15 & 0.2 \\
\multicolumn{5}{l}{Dangling Band Donor} \\
\quad Electron Thermal Cross Section & cm${}^{2}$ & $3\times 10^{-14}$ & $3\times 10^{-14}$ & $3\times 10^{-14}$ \\
\quad Hole Thermal Cross Section & cm${}^{2}$ & $3\times 10^{-15}$ & $3\times 10^{-15}$ & $3\times 10^{-15}$ \\
\quad Total Trap Density & cm${}^{-3}$ & $10^{16}$ & $10^{16}$ & $10^{16}$ \\ 
\quad Energy of Distribution & eV & 1.1 & 0.9 & 0.45 \\
\quad Characteristic Energy & eV & 0.2 & 0.15 & 0.2 \\ \hline 
\end{tabular}
\end{center}						
\end{table*}
\subsection{Conical Substrate}
After the radiation of the first photons to the inner surface of the cone, a chain of radiation and reflection begins. Depending on the vertex angle of the cone and the location of the first beam, the number of times a photon hits the cell surface will vary from one to several times. In the same way, the chances of photon absorbing increases. In this concept, the possibility of recycling the photons exited out of the film can improve the cell's efficiency. These radiations take place at different angles. Trigonometric and geometric calculations give the following relations for the angle of each radiation:
\begin{equation} \label{equ2} 
a\buildrel\Delta\over= \frac{H}{R} =\cot \left(\frac{\theta _{v} }{2} \right) 
\end{equation} 
\begin{equation} \label{equ3} 
m_{k} \buildrel\Delta\over= \tan \left(\varphi _{k} \right) 
\end{equation} 
\begin{equation} \label{equ4} 
x_{k+1} =\frac{a-m_{k} }{a+m_{k} } x_{k}  
\end{equation} 
\begin{equation} \label{equ5} 
m_{k+1} =\frac{2a+(1-a^{2} )m_{k} }{1-a^{2} -2am_{k} }  
\end{equation} 
\begin{equation} \label{equ6} 
\alpha _{k+1} =2\theta _{v} -\alpha _{k}  
\end{equation} 
Where $\theta _{v}$ is the vertex angle of the cone, and $x_{k}$, $\varphi _{k}$, $\alpha _{k}$ are respectively the incident light's location, angle to the horizon, and angle to the surface.
\subsection{Paraboloid substrate}
Only one or two incidents take place for paraboloid cells. The paraboloid can be described using the following equation:
\begin{equation} \label{equ7} 
y=\frac{H}{R^{2}} x^{2} =\frac{a}{R} x^{2}  
\end{equation} 
If each vertical incidence has a distance more than a critical value to the paraboloid axis, it will have another incident with the inner surface of the paraboloid after reflection and passing through the focal point and it will vertically exit after the second reflection. However, if the distance is lower than the mentioned value, the reflected beam will exit without the re-encountering the surface of the paraboloid, hence the total encounter of each photon to the inner surface of the paraboloid will be 2 times maximum. The critical distance, $R_{C}$,  can be derived as:
\begin{equation} \label{equ8} 
R_{C} =\frac{R^{3}}{4H^{2}} =\frac{R}{4a^{2} }  
\end{equation} 
The angle and length from the origin of these two incidents are obtained via the following equation:
\begin{equation} \label{equ9} 
\varphi _{1} =\tan^{-1} \left(\frac{R^{2} }{2Hx_{0} } \right)=\tan ^{-1} \left(\frac{R}{2ax_{0} } \right) 
\end{equation} 
\begin{equation} \label{equ10} 
\varphi _{2} =90-\varphi _{1}  
\end{equation} 
\begin{equation} \label{equ11} 
x_{2} =-\frac{R^{4} }{4H^{2} x_{1} } =-\frac{R^{2} }{4a^{2} x_{1} }  
\end{equation} 
\begin{figure}
	\begin{center}
		\resizebox{0.45\textwidth}{!}
		{\includegraphics{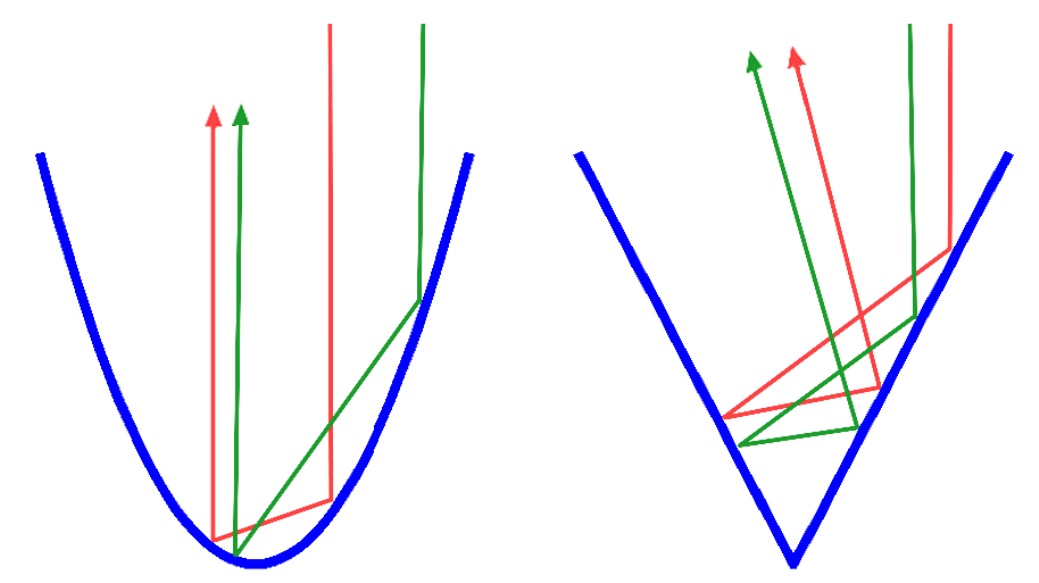} }
		\caption{\label{fig5}Cross sections of paraboloid and cone shapes against two sample rays.}
	\end{center}
\end{figure}
\section{Results}
\begin{figure}
	\begin{center}
		\resizebox{0.75\textwidth}{!}
		{\includegraphics{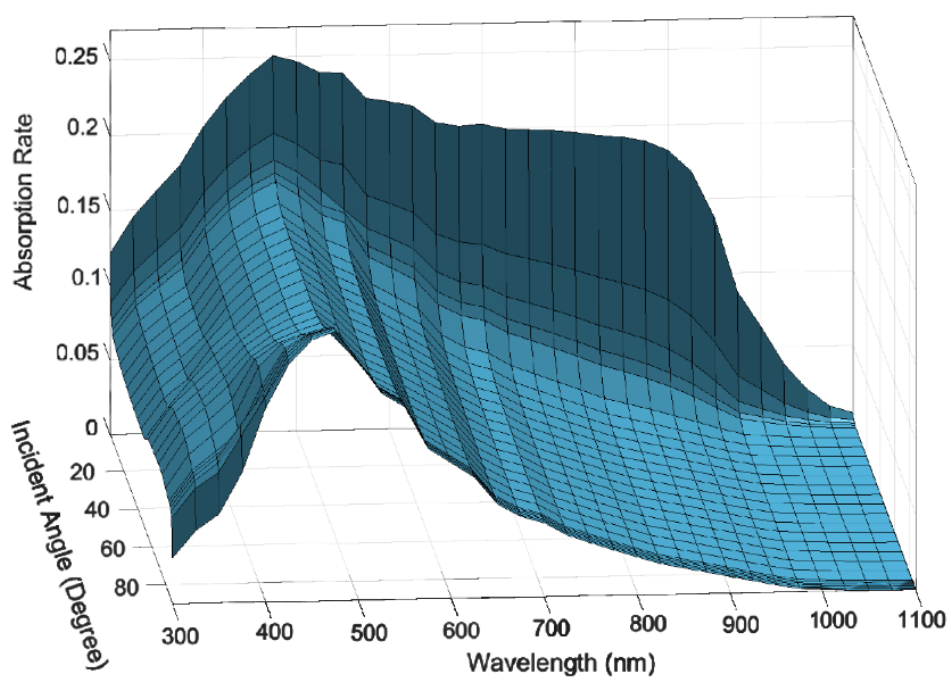} }
		\caption{\label{fig6}Absorption rate of base cell versus photons wavelength and incident angle.}
	\end{center}
\end{figure}

\begin{figure} 
	\begin{center}
		\resizebox{0.45\textwidth}{!}
		{\includegraphics{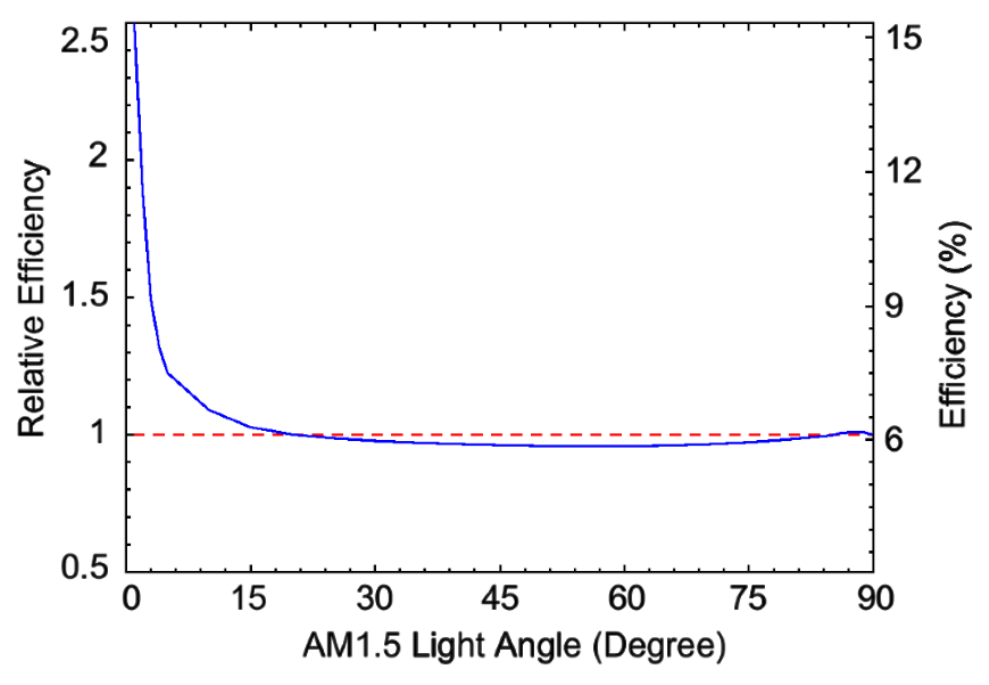} }
		\caption{\label{fig7}Efficiency of the base cell versus AM1.5 light angle.}
	\end{center}
\end{figure}

\begin{figure}
	\begin{center}
		\resizebox{0.45\textwidth}{!}
		{\includegraphics{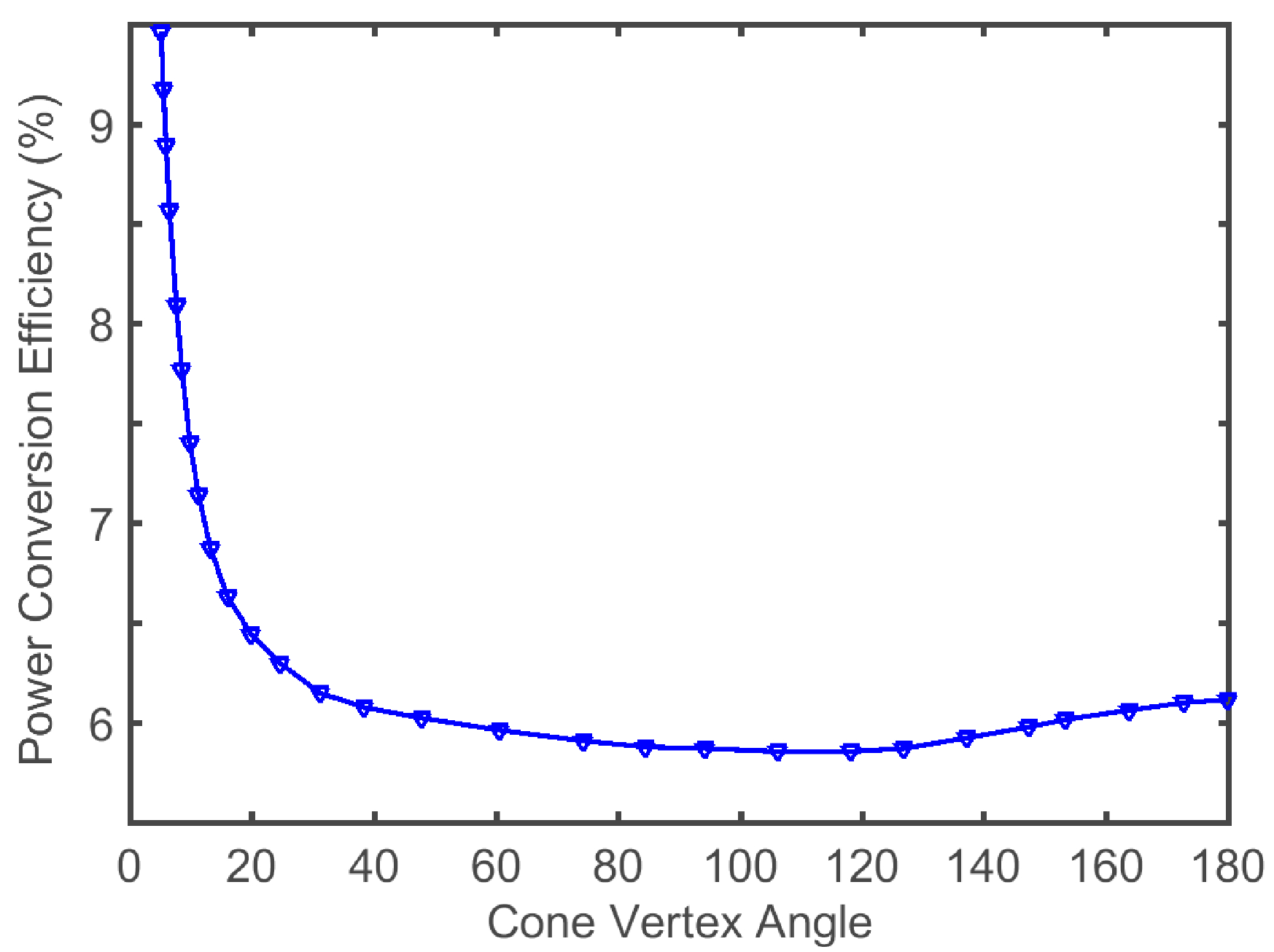} }
		\caption{\label{fig8}Efficiency of the cone-shaped cell versus the cone's vertex angle.}
	\end{center}
\end{figure}

\begin{figure}
	\begin{center}
		\resizebox{0.45\textwidth}{!}
		{\includegraphics{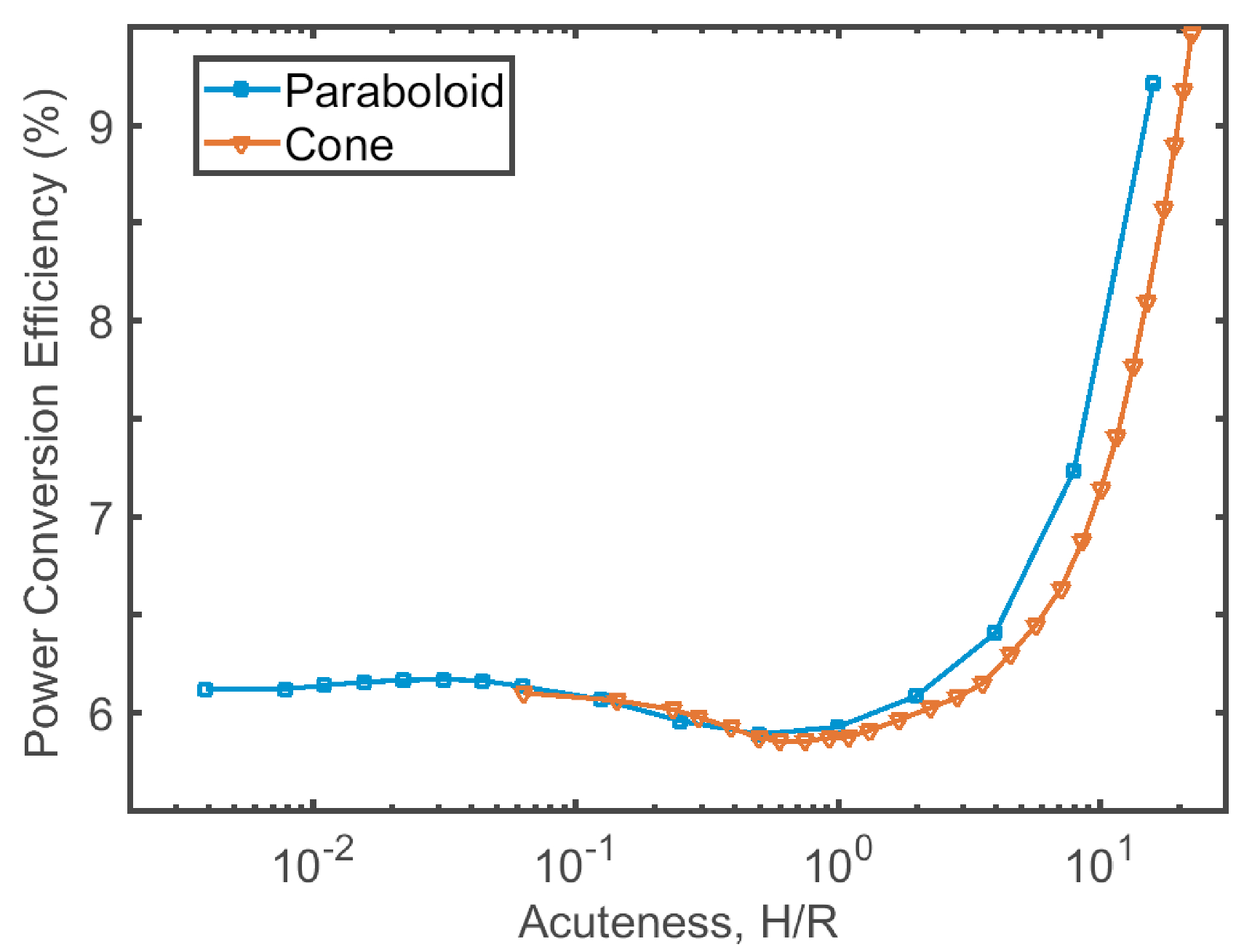} }
		\caption{\label{fig9}Efficiency of two studied cell shapes versus their acuteness'.}
	\end{center}
\end{figure}

\begin{figure}
	\begin{center}
		\resizebox{0.45\textwidth}{!}
		{\includegraphics{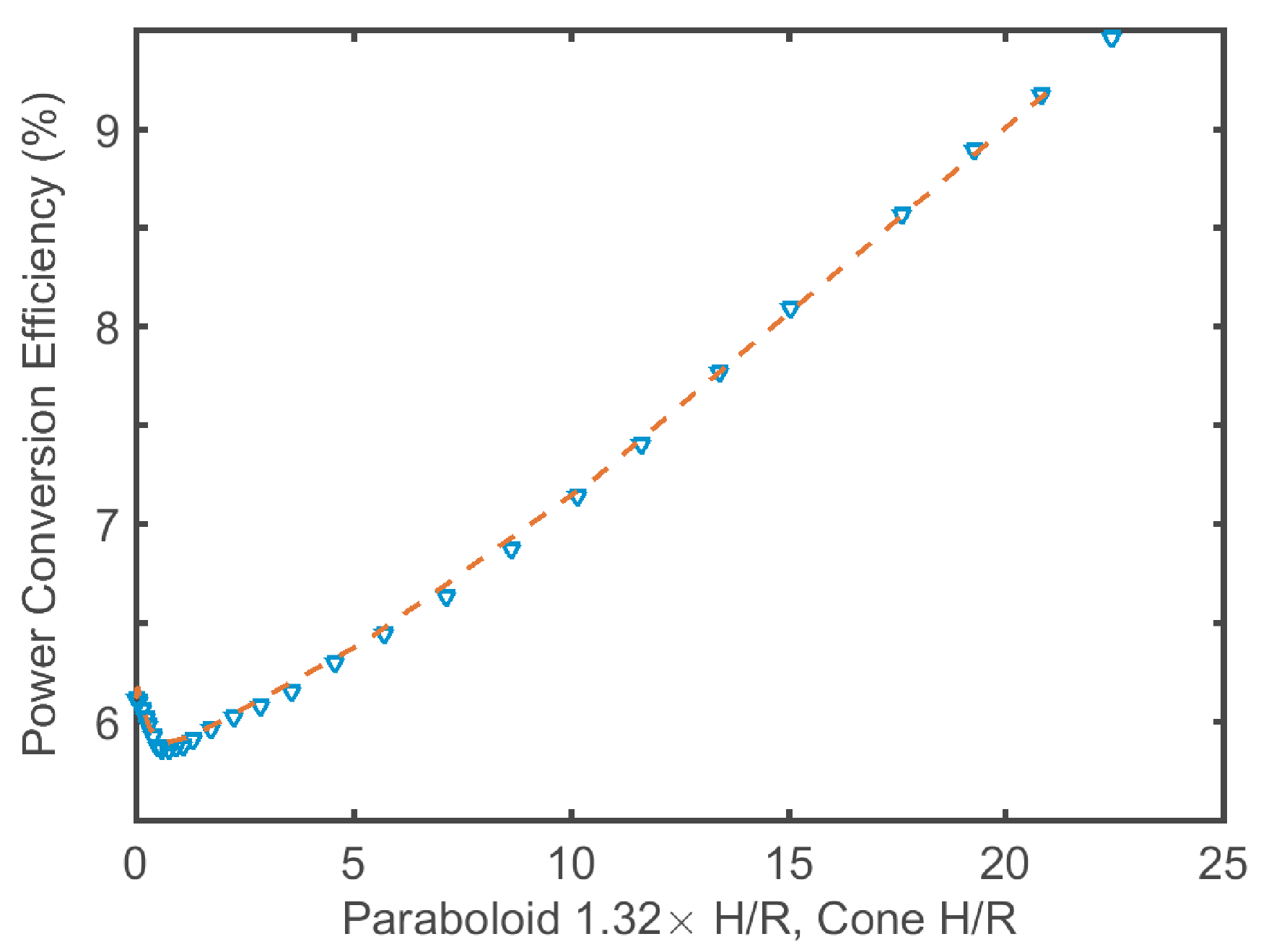} }
		\caption{\label{fig10}Efficiency of two studied cell shapes versus acuteness in different scales.}
	\end{center}
\end{figure}

After parametric simulations of the base cell discussed in the simulation section, the wavelength, incident angle and absorption rate database is filled. The surface in Fig. \ref{fig6} shows the graphical view of the absorb rate extracted from this database.
By combining the wavelengths of the photons according to the AM1.5 spectra and applying the data of this database, the cell efficiency is obtained in terms of incident light angle, which is shown in Fig. \ref{fig7}. The incident angle is calculated relative to the cell surface. As shown in this figure, when the incident angle is reduced to less than 16\degr, the cell's efficiency increases compared to the vertical incident, and at an angle of 1\degr, it becomes 2.5 fold. It should be noted that radiation with a very small angle leads to many technological problems.\\
As shown in Fig. \ref{fig5}, it is assumed that the first ray emitted from the sun is parallel to the main axis of the cone/paraboloid. This assumption is the base of the current study and non-orthogonal ray emission can be investigated in the future works.\\
By applying the AM1.5 light and following the radiation-reflection-exit chain explained in equations (\ref{equ2})--(\ref{equ6}) and extracting the necessary data from the database, the efficiency of the solar cell with the conical substrate in terms of the vertex angle of the cone is obtained as shown in Fig. \ref{fig8}. As it can be seen, for vertex angles less than 30\degr, the cell's efficiency increases by decreasing the vertex angle, and at an angle of 2\degr, it reaches 9.4\%, that is, about 53\% better than the flat cell. it should be considered that making a conical cell with a very steep angle is quite challenging.
In the case of the paraboloid, similar calculations and simulations were carried out, and, as a result, the efficiency of the cell with paraboloid substrate was calculated. In order to have better criteria than complexity and difficulty, and comparison with conical cells, the efficiency of both cells is plotted in Fig. \ref{fig9} in terms of the acuteness. As shown, raising the acuteness of conic (paraboloid) cell for an acuteness greater than 3(2), increases efficiency. However, nearly 50\% improvement in the efficiency in comparison to the flat cell is observed when the acuteness reaches 20(14).
This significant increase in power conversion efficiency can be observed in some similar researches such as \cite{RN13b,RN13d}. Nonetheless, the effect of V-shaped cell's opening angle variations in the recent references was investigated from 180\degr (flat) to 30\degr. Under this condition an increase in efficiency was reported from 2.2\% to 3.5\% and 1.3\% to 2.1\% in \cite{RN13b} and \cite{RN13d}, respectively.\\
It should be highlighted that constructing very acute cells also has technological limitations. Fig. \ref{fig10} shows two curves: the cone-shaped cell efficiency in terms of its acuteness and the paraboloid cell efficacy in terms of 1.32$\times$ its acuteness. These two curves fit well together. The purpose of this comparison is to prove that for an identical opening radius, the cone-shaped cell has a 1.4$\times$ height in comparison to paraboloid cell with similar efficiency. The ability to extract cones from cut flat cells is an advantage of conical substrates.\\
In the present approach, the optical part of simulation is assumed to be independent of the electrical part. Generally, just the optical effects caused by recycling of the reflected photons has been considered to calculate the device efficiency and the probable interplay between the optical and electrical effects is neglected. However, if the sheet resistance of ITO layer is small enough, and/or an appropriate metal grids is used, it is possible to ignore the parasitic effects raised by the excess series resistance of large dimension and/or very acute cells.
\section{Conclusions}
First, a method for calculating the efficiency of a thin-film solar cell with a non-flat substrate was presented by simulating a base cell along with auxiliary geometric computing. In this method, a database on the absorption of the flat cells is formed in terms of radiation angle and photon wavelength. Then, geometric and trigonometric calculations are used to track the radiation-reflection chain until photon exits from the cell, and the chance of absorbing each photon at the end of the chain is computed. By summing up the results for each spectrum and the desired intensity of light, including AM1.5, the efficiency of each non-flat cell will be calculated.\\
Using the proposed method, the efficiency of thin-film cell was calculated with two non-flat paraboloid and conical substrate. It was observed that if in these two cells, the ratio of height to radius was large enough, it would be possible to trap more photons than the flat cells and create better efficiency. Radius and substrate height can be relatively large, up to a centimeter, so that they can be easily constructed in a common deposition system.
\section{Acknowledgment}
This study was supported by a research grant from Islamic Azad University, Mahshahr Branch, Mahshahr, Iran.




\section{References}

 \bibliographystyle{elsarticle-num} 





\end{document}